\documentclass{appolb}
\usepackage{graphicx}
\usepackage{comment}
\usepackage{cite}

\begin{document}
\title{Determination of the polarization in the $\vec{p}p\to pp$ reaction with WASA-at-COSY detector%
}
\author{I.~Sch{\"a}tti-Ozerianska, P.~Moskal, M.~Zieli{\'n}ski
\address{The Marian Smoluchowski Institute of Physics, Jagiellonian University\\ \L{}ojasiewicza 11, 30-348 Krak\'{o}w, Poland}\\
and\\
\address{Institut f\"{u}r Kernphysik, (IKP), Forschungszentrum J\"{u}lich\\ Wilhelm-Johnen-Stra{\ss}e 52428 J\"{u}lich, Germany} 
\\
and\\
\address{the WASA-at-COSY Collaboration}
}
\maketitle
\begin{abstract}
The dynamics of $\eta$-meson production and the interaction of $\eta$-mesons with nucleons can be studied using the
$\vec{p}p\to pp\eta$ reaction via measurements of the analyzing power $A_{y}$.
To this end, we have performed a measurement of the $\vec{p}p\to pp\eta$ reaction using the  
 WASA-at-COSY detector, which provides large acceptance and is $\varphi$-symmetric. The experiment was carried out for beam momenta of $2026$~MeV/c and $2188$~MeV/c. 
In this article we present a method and results for the determination of the degree of the beam polarization.
\end{abstract}
  
\section{Introduction}

Despite a number of experiments~\cite{chiavassa94,calen96,calen97,hibou98,smyrski00,bergdolt93,abdelbary03,
ETA-PRC-Moskal,ETA-EPJ-Moskal,ETA-Petren,pn-deta-Calen,pn-pneta-Calen,pn-pneta-Moskal} 
and theoretical approaches~\cite{ACTA-Colin,nakayama02,faldt01,germond90,laget91,moalem96,vetter91,alvaredo94,batinic97} still an open question remains
about the mechanism of the $\eta$-meson production in nucleon-nucleon collisions. Previous studies concluded that 
the $\eta$-meson production in $pp$ and $pn$ reactions occurs predominantly via the $N^{*}(1535)$ resonance 
and that the proton-$\eta$ interaction is much 
larger than in the case of proton-$\pi^0$, and proton-$\eta'$~\cite{moskal00,review,Czerw1,Czerw2,Nanova}.
For an unambiguous understanding of the $\eta$ production process, relative magnitudes 
from the partial wave contributions must be well established. The incompatibility between experiment and theory may be 
resolved by taking into account higher partial waves and additional baryon resonances~\cite{nakayama03}.
 This kind of studies can be performed by measuring analyzing power $A_{y}$ of the $\eta$-meson.
Up to now, measurements of $A_y$ in the $\vec{p}p\to pp\eta$ reaction were conducted 
by the COSY-11 and DISTO collaborations~\cite{aycosy11,ETA-Ay-EPJ-Winter,ETA-Ay-PL-Winter,ETA-Ay-Balestra}.
However, these previous experiments suffered from low statistics and limited geometrical acceptance~\cite{Brauksiepe,C11Klaja}.
In order to determine $A_y$ more accurately, the axially symmetric WASA detector~\cite{HHAdam} 
and the vertically polarized proton beam of COSY have been used to collect a high statistics data sample.
An experiment was conducted for beam momenta 2026 MeV/c and 2188 MeV/c which 
correspond to excess energies of 15~MeV and 72~MeV, respectively~\cite{spin2010,Prop10}.
The trigger was set to select events corresponding to the $\vec{p}p\to pp\eta$ and $\vec{p}p\to pp$ reactions. Proton-proton elastic scattering 
provides information for the determination of the polarization and the luminosity.
Additionally, to control the effects caused by potential asymmetries in
the detector setup the spin orientation was flipped from cycle to cycle. 

\section{Extraction of the $pp \to pp$ reaction}
\begin{figure}[!h!]
\begin{center}
\includegraphics[width=0.6\linewidth]{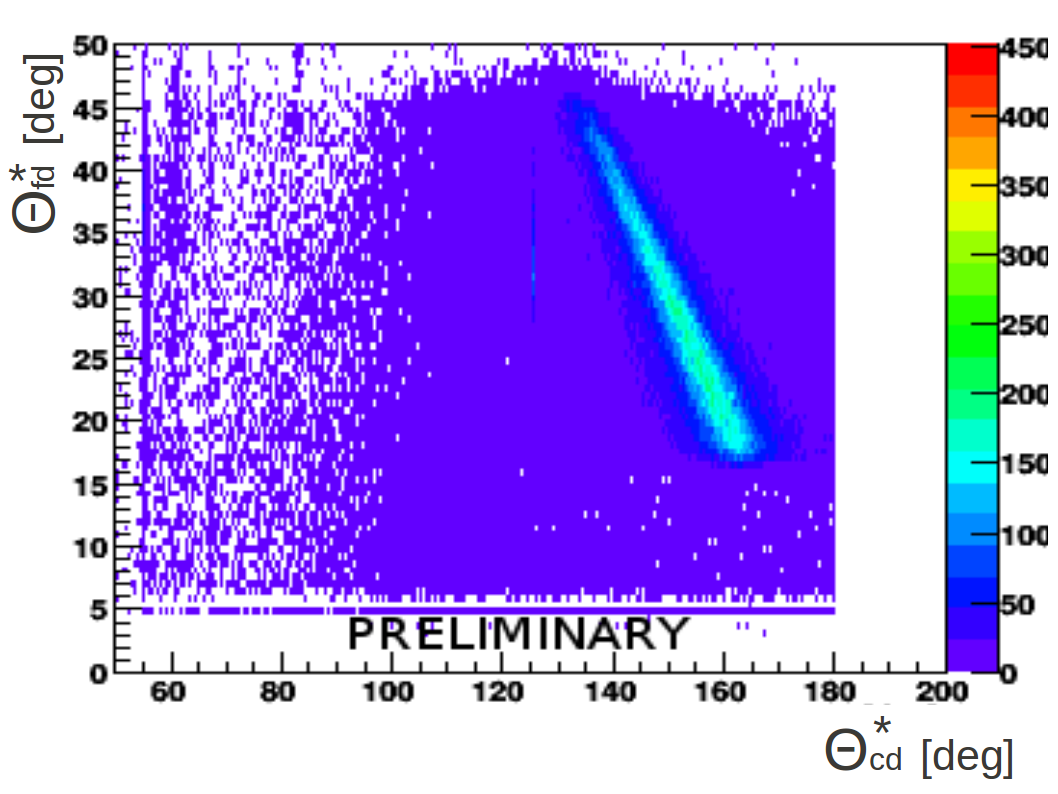}
\caption{Example of the $\theta^*_{FD}$ vs $\theta^*_{CD}$ distribution.}
\label{theta1}
\end{center}
 \end{figure}

The first step in the study of the asymetry $A_y$ is the analysis of the reaction $\vec{p}p\to pp$.
Based on the elastic scattering reaction we have calculated the polarization degree for the whole data sample. 
In the $\vec{p}p \to pp$ reaction one of the protons is registered in the Forward Detector, with geometrical
acceptance of polar angle $\theta_{FD}$ from $3^{\circ}$ to $18^{\circ}$ and the second is registered in the  Central Detector. 
By utilizing $\bigtriangleup$E -E method in the Forward Range Hodoscope and the angular correlation of the outgoing protons we have selected events corresponding 
to the proton-proton elastic scattering. For further calculations we have chosen only two $\theta^*_{FD}$ ranges in the centre of mass 
[$30^{\circ}-34^{\circ}$] and [$34^{\circ}-38^{\circ}$], for which $A_y$ is available from EDDA database~\cite{Altmeier2000} and for which the number of event statistic was sufficient.
In order to estimate the background, for each bin in $\theta^*_{FD}$, the $\theta^*_{CD}$ distribution was fitted separately with a fifth-order 
polynomial, excluding in the fit the range from 130$^\circ$ to 160$^\circ$, as shown in Fig.~\ref{theta1}.

\begin{figure}[h!]
\begin{center}
\includegraphics[width=0.8\linewidth]{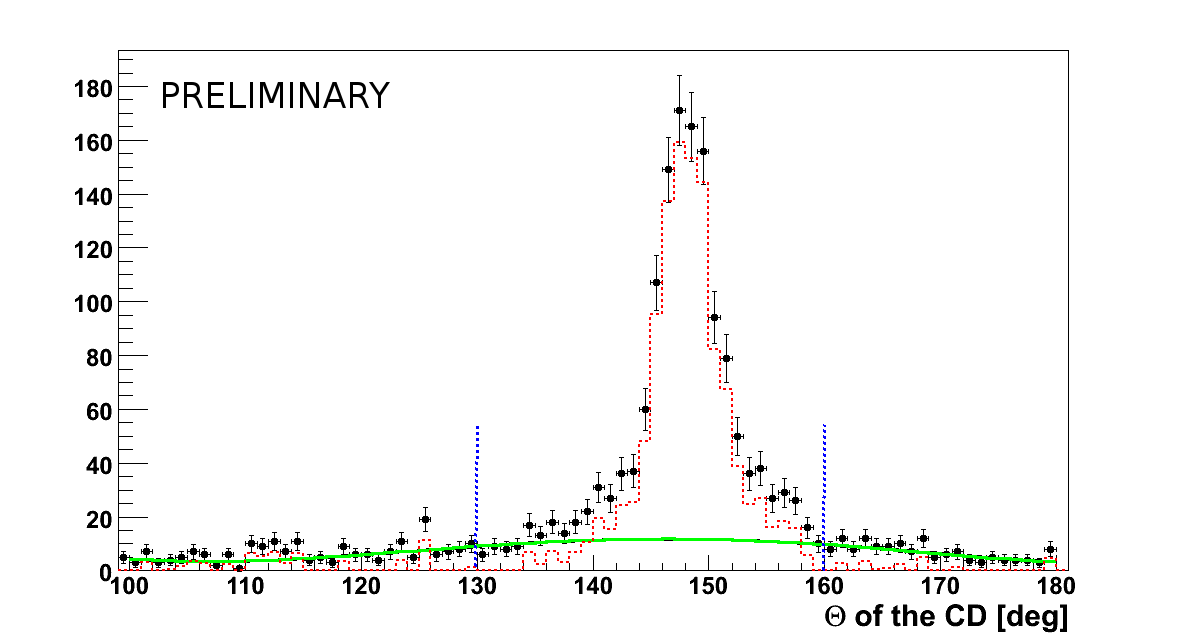}
\end{center}
\caption{The distribution of $\theta^*_{FD}$ for $\theta^*\in [30^{\circ},34^{\circ}]$, $\varphi\in [20^{\circ},30^{\circ}]$ and spin mode. Points denote data. The green line corresponds to the fitted background, and the dotted red histogram shows the data after subtraction of the background.}
\label{theta2}
\end{figure}

Subsequently in order to estimate a background each separate $\theta^*_{CD}$ distribution was fitted 
with the fifth polynomial excluding in the fit the range from 130$^\circ$ to 160$^\circ$.
Example of such distribution can be seen in Fig.\ref{theta2}

\begin{figure}[h!]
\includegraphics[width=0.55\linewidth]{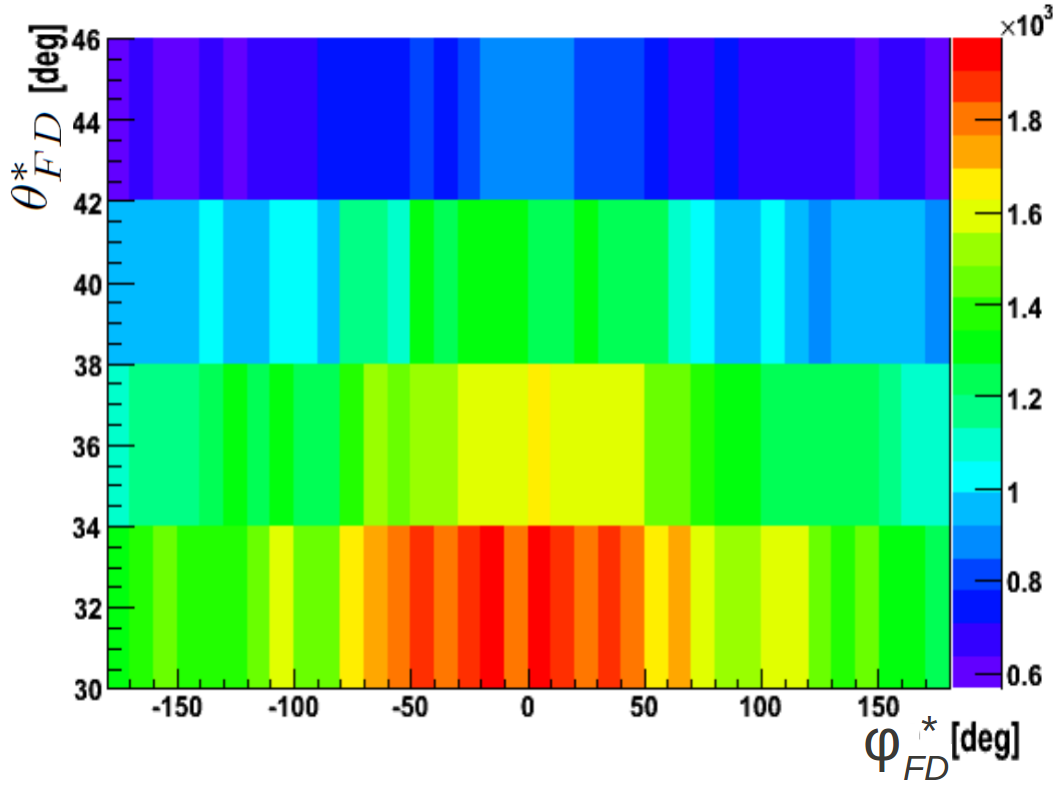}
\includegraphics[width=0.55\linewidth]{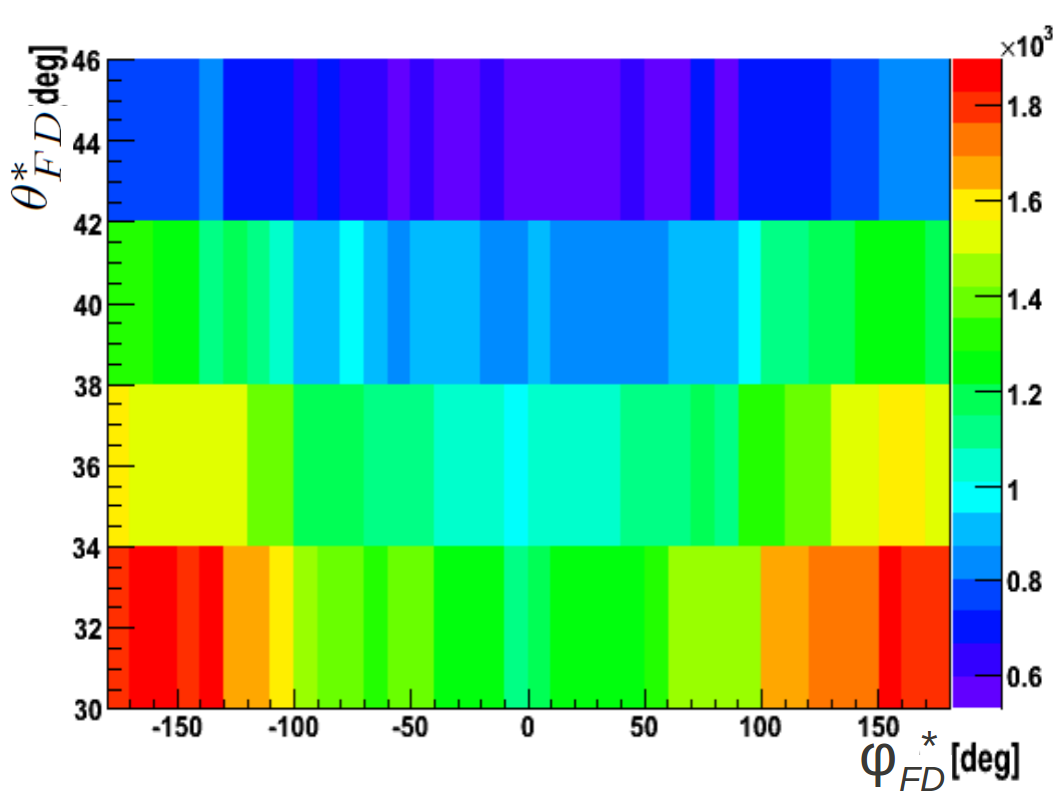}\\
\caption{Example of the angular distribution for the elastic scattered events. Left: spin up mode. Right: spin down mode.}
\label{theta3}
 \end{figure}
\section{Polarization determination}

The polarization was determined from elastic scattering events by the following equation: 
\begin{equation}
\frac{N^{exp}(\theta,\varphi)-N^{exp}(\theta,\varphi+180)}{N^{exp}(\theta,\varphi)+N^{exp}(\theta,\varphi+180)} = P\cdot cos\varphi \cdot A_{y}(\theta)
\label{eq:coef}
\end{equation}
where $N^{exp}(\theta,\varphi)$ and $N^{exp}(\theta,\varphi+180)$ correspond to a given angular bin after correction for the acceptance, and P denotes the polarization.
$A_{y}(\theta)$ was calculated based on the results of the EDDA experiment~\cite{Altmeier2000}. 
The procedure is described in detail elsewhere~\cite{Hodana:2013cga}.

\begin{figure}[h!]
\includegraphics[width=0.47\linewidth]{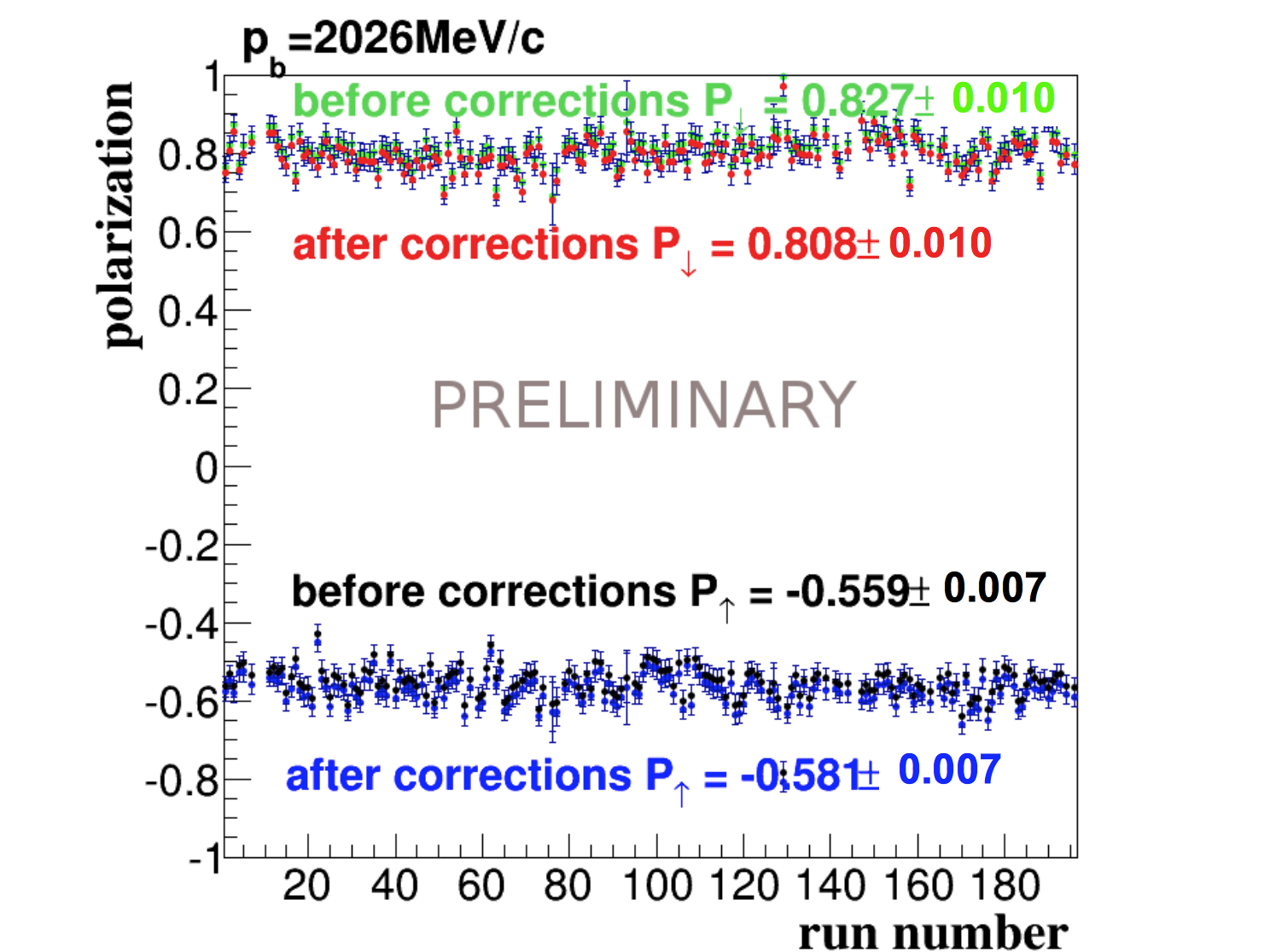}
\includegraphics[width=0.47\linewidth]{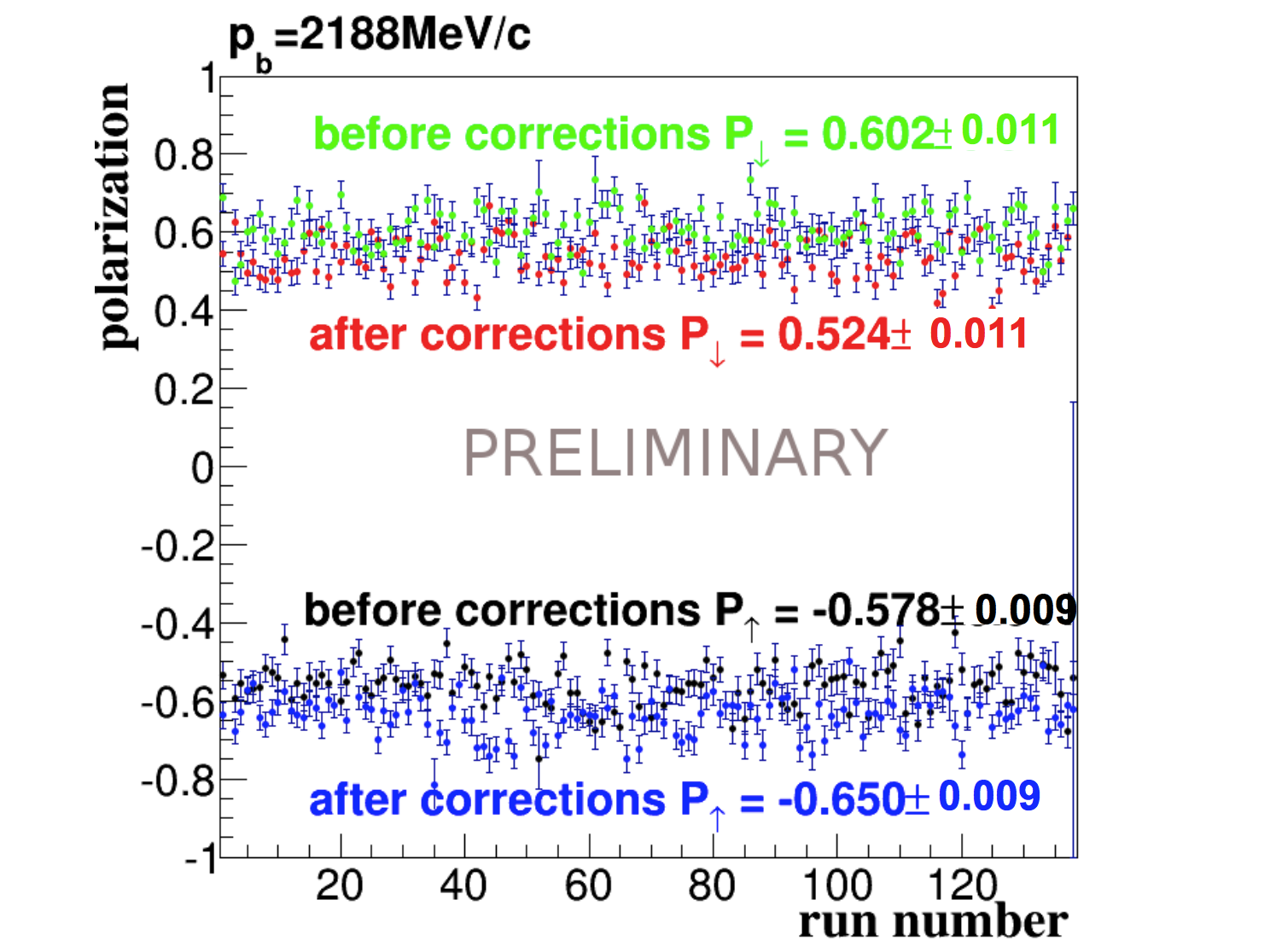}\\
\begin{center}
\includegraphics[width=0.47\linewidth]{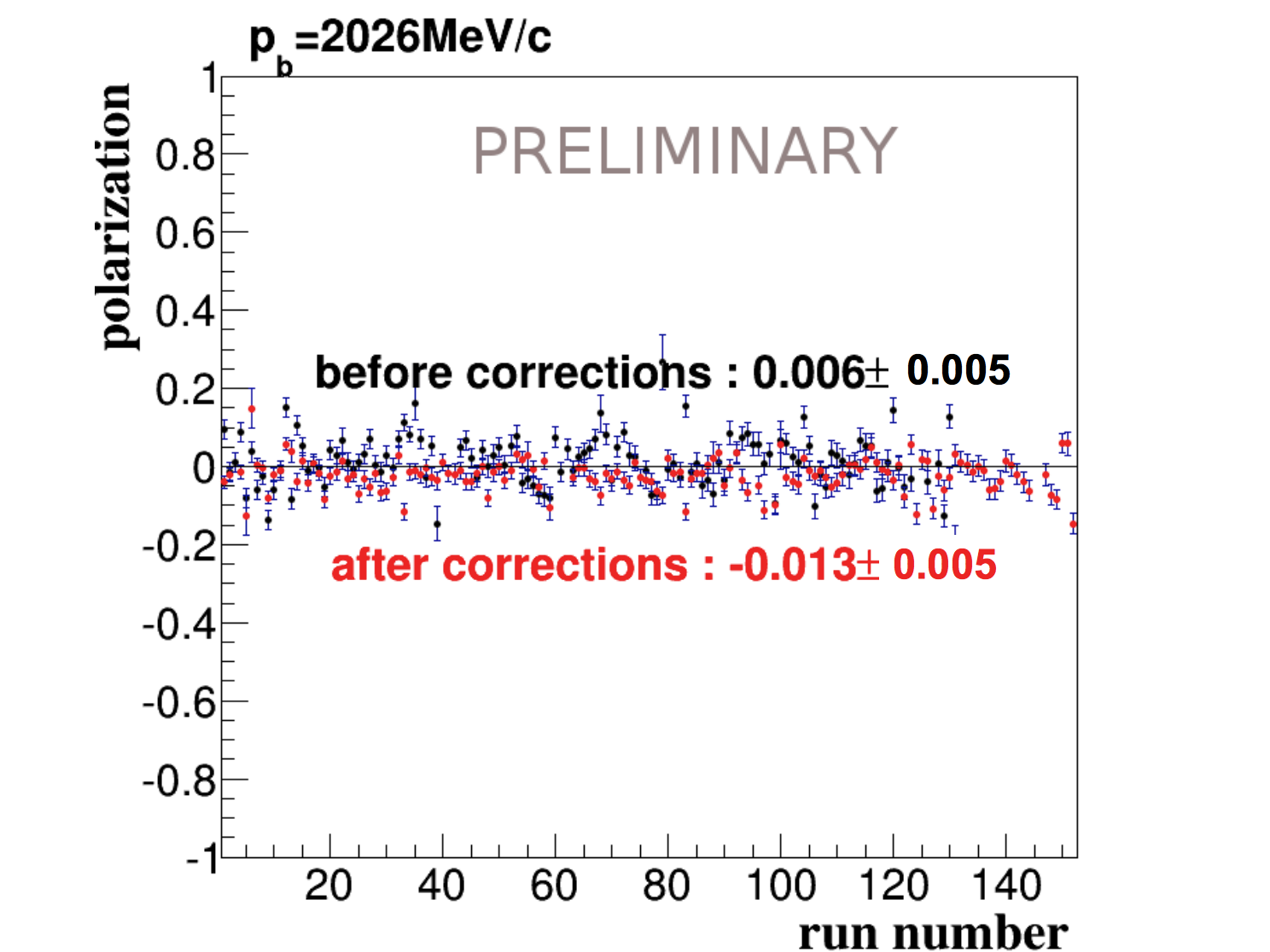}
\end{center}
\caption{Distributions of polarization as a function of run number for polarized (upper panel) and unpolarized (lower panel) data. 
Results for both polarization modes are shown.}
\label{f:Pol}
\end{figure}

The polarization determined from all experimental data is shown in Fig.~\ref{f:Pol}. 
The upper panel shows the results for measurements with a polarized beam obtained for both studied beam momenta.
The lower panel presents the polarization of the control sample, which has been measured with an unpolarized beam.
In the case of the control sample, the measured polarization is consistent with zero, within the statistical uncertainty.
Fig. 4 shows the results before and after the correction for shifts of the center the interaction region between the beam and the target~\cite{Hodana:2013cga,hi_hod}.

 The average polarization of all acquired data was calculated analyzing all events together. 
The results are shown in Tab.\ref{taballpol}.
\begin{table}[h!]
\begin{small}
\begin{center}
\begin{tabular}{|c|c|c|}\hline
$p_{b}[GeV/c]$ & spin mode & polarization \\
\hline
  2026 &down & 0.793$\pm$0.010\\
       &up   & -0.577$\pm$0.007\\
\hline
  2188 &down & 0.537$\pm$0.009\\
       &up   & -0.635$\pm$0.011\\
\hline
 2026 &unpolarized& -0.012$\pm$0.005\\
\hline
\end{tabular}
\end{center}
\begin{center}
\caption{Preliminary polarization achieved in the WASA-at-COSY experiment conducted in the year 2010\label{taballpol}.}
\end{center}
\end{small}
\end{table}

\newpage
\section{Summary}
The polarization of elastic scattering events in the reaction $\vec{p}p\to pp$ has been presented. 
The polarization differs between the spin modes. However, it is stable between runs for the whole period of data taking. Additionally, 
the value of the polarization for an unpolarized beam has been measured to be consistent with zero, as expected.
A high degree of polarization for both beam momenta 
gives a very good perspective for the determination of the analyzing power for the $\vec{p}p\to pp\eta$ reaction. 
\subsection{Acknowledgements}
We acknowledge support 
by the Polish National Science Center through grant No. 2011/03/B/ST2/01847, 
by the FFE grants of the Research Center J\"{u}lich.

\end{document}